\documentclass[reprint,aps,pra,floatfix,longbibliography]{revtex4-1}
\usepackage{float}
\usepackage{graphicx}
\usepackage{braket}
\usepackage{amsmath,mathtools}
\usepackage{color}

\newcommand{\p}[2]{\frac{\partial #1}{\partial #2} }
\newcommand{\etal}{\textit{et al.} }
\newcommand{\D}{\, \mathrm{d}}

\newcommand{\dbra}[1]{\Braket{\Braket{#1}}}
\newcommand{\eqreft}[1]{Eq.~\eqref{#1}}

\usepackage{hyperref}
\hypersetup{colorlinks=true,citecolor=blue,linkcolor=blue,urlcolor=blue}

\begin{document}
\title{Floquet analysis of time-averaged trapping potentials}
\author{Oliver A. D. Sandberg}
\affiliation{ARC Centre of Excellence in Future Low-Energy Electronics Technologies (FLEET), School of Mathematics and Physics, The University of Queensland, Brisbane, Queensland 4072, Australia}
\author{Matthew T. Reeves}
\affiliation{ARC Centre of Excellence in Future Low-Energy Electronics Technologies (FLEET), School of Mathematics and Physics, The University of Queensland, Brisbane, Queensland 4072, Australia}
\author{Matthew J. Davis}
\email{mdavis@physics.uq.edu.au}
\affiliation{ARC Centre of Excellence in Future Low-Energy Electronics Technologies (FLEET), School of Mathematics and Physics, The University of Queensland, Brisbane, Queensland 4072, Australia}

\begin{abstract}
	Time-averaged trapping potentials have played an important role in the development of the field of ultracold atoms.  
	Despite their widespread application, there is not yet a complete understanding of when a system can be considered time-averaged.
	Here we use Floquet theory to analyse the lowest energy state of time-periodic trapping potentials, and characterise the transition from a localised state in a slowly moving trap to a delocalised state in a rapidly oscillating time-averaged potential. 
 We investigate how the driving parameters affect the density and phase of the Floquet ground state, 
	and provide a quantitative measure of the degree to which they can be considered time-averaged.
 We study a number of simple representative systems, and comment on the features affecting the experimental realisation of time-averaged trapping potentials.
\end{abstract}
\maketitle
\section{Introduction}
Ultracold atomic gases provide a versatile testing ground for the study of quantum many-body physics.
The ability to precisely control experimental 
conditions such as trapping potentials and interaction parameters allows them to be used as 
 a toolbox for designer matter, and their relative simplicity often allows for direct comparison with theory \cite{Goldman2016}. 
The high degree of control and cleanliness can allow the investigation of novel phases of matter and novel 
far-from-equilibrium phenomena which are often not accessible in solid-state systems \cite{Manzoni2017}. 
 
The flexible control and engineering of trap geometries has been an important feature in cold atom research, opening intriguing possibilities for quantum simulation \cite{Bloch2012}, quantum computation \cite{Zipkes2010} and the creation of exotic states of matter \cite{Greiner2003, Li2017}.
One approach to trap design has been the use of time-averaging: moving a trapping potential at a frequency greater than the atoms can respond to kinematically so that the effective trap is stationary with respect to the characteristic time scale of their evolution.
Examples include the Time Orbiting Potential (TOP) trap which Petrich \etal used in the original quest for BEC \cite{Petrich1995,Han1998}, and rapidly-scanned optical dipole traps \cite{Friedman2000,Milner2001,Rudy2009,Schnelle2008,Henderson2009,Gildemeister2010,Sherlock2011,Bell2016,Bell2018}.  In a similar fashion, experiments have utilised shaken optical lattices to modify effective tunnelling rates between lattice sites \cite{Lignier2007,Sias2008,Creffield2010}.
Experiments generally drive the trapping potential as fast as is technically possible to ensure they are in the time-averaged regime, following a rough guideline of $\Omega \gg \omega$ --- that the driving frequency $\Omega$ is much faster than the frequency of the trap $\omega$.
This raises the question: what precisely are the conditions for which the time-averaging approximation is effective?

It is important to note  that the effective static trap in the time-averaged limit is merely an approximation, and the system will still exhibit some dynamical features due to the driving.  Experimentally this often manifests as a reduced trap lifetime, and/or an inherent heating rate of the atoms \cite{Gildemeister2010}.
It is therefore important to understand in greater detail how these features manifest, and correspondingly how they might be minimised. 
Some theoretical work has been done in investigating the underlying dynamical effects induced by the drive, including investigations of micromotion \cite{Muller2000, Challis2004}, but a study of the full transition from states localised in a static trap to becoming delocalised in the time-averaged limit has yet to be undertaken.  

A natural approach to address these issues is Floquet theory, which provides a convenient basis in which to investigate time-periodic systems.  
The Floquet framework has been used in investigations of topological states \cite{Rechtsman2013,Mikami2016,Swingle2005}, the engineering of artificial gauge fields \cite{Kuwahara2016,Dalibard2011}, synthetic magnetic fields \cite{Creffield2016}, spin-orbit couplings \cite{Wu2006,Li2017a,Sun2018} and artificial atoms~\cite{Deng2015}.

Here we apply a Floquet analysis to periodically-driven trapping potentials.  We examine the nature of the transition from slow driving, where the lowest energy states of the systems are localised and adiabatically follow the moving potential, to fast driving, where the lowest energy states are delocalised in the time-averaged potential. 
The precise way that the system couples to the drive determines how the localised to delocalised transition occurs.
We provide a quantitative measure of how well the system approximates the time-averaged limit and additionally derive analytical results which give insights into the time-averaged transition.  

This paper is organised as follows.  In Sec.~\ref{sec:floquet} we provide a summary of Floquet theory and a description of our numerical approach.   In Sec.~\ref{sec:ring} we analyse a ring potential formed by a time-averaged attractive Gaussian trap. 
This system is both experimentally relevant \cite{Henderson2009,Sherlock2011,Bell2016,Bell2018}, and relatively simple, with a Galilean transformation allowing analysis in a stationary frame.
In Sec.~\ref{sec:nonGalilean}, we apply our analysis to three other representative one-dimensional potentials, demonstrating key features of time-averaged systems that are important for the design and analysis of experimental setups.
We highlight the different kinds of resonances that emerge for different trapping potentials.
In Sec.~\ref{sec:harmonic} and Sec.~\ref{sec:abs}, we study systems which are harmonic in the time-averaged limit, so they display a collective resonance.
In Sec.~\ref{sec:quartic}, we study an anharmonic system, for which a collective resonance does not occur. 
In Sec.~\ref{sec:abs} and Sec.~\ref{sec:quartic} we see the emergence of so-called ``photon'' resonances, which are responsible for uncontrollable heating in experimental systems \cite{Eckardt2015}.  Finally, we conclude in Sec.~\ref{sec:conclusions}.

\section{Floquet theory}
\label{sec:floquet}
In order that this paper is self-contained we provide a brief overview of Floquet theory.  For a more complete description we refer the reader to Refs.~\cite{Grifoni1998c,Reichl1992}.

For systems with a periodic time-dependence a stationary eigenbasis does not exist.
Instead, an alternative is the \emph{stroboscopic} basis in which the states are stationary only when sampled in integer multiples of the driving period, $T$ (frequency $\Omega = 2\pi/T$). 
The technique of Floquet analysis combines the usual Hilbert space of square-integrable functions, $\mathcal{R}$,  with the Hilbert space of all time-periodic functions $\mathcal{T}$ to form the composite Hilbert space $\mathcal{R} \otimes \mathcal{T}$.  
This composite space has norm \cite{Sambe1973}
\begin{align}
	\dbra{a(t)|b(t)} &\equiv \frac{1}{T}\int_{0}^{T} \int a^*(x,t) b(x,t) \D x \D t, \\
	&= \frac{1}{T}\int_{0}^{T}\Braket{a(t)|b(t)}\D t
	\label{timenorm}
\end{align}
which is a natural combination of the well-known norms of the constituent Hilbert spaces $\mathcal{R}$ and $\mathcal{T}$.

For a periodic Hamiltonian $H(t+T) =H(t)$ with period $T$, Floquet's theorem \cite{Floquet1883} implies that there exist so-called Floquet-state solutions to the Schr\"odinger equation 
\begin{align}
i \hbar \p{}{t} \Psi(x,t) =  H(x,t) \Psi(x,t), \label{Schro}
\end{align}
of the form,
\begin{align}
\Psi_\alpha(x,t) = e^{-i \varepsilon_\alpha t/\hbar } \Phi_\alpha(x,t), \label{floquetsolution}
\end{align}
where $\Phi_{\alpha}(x,t)$ is a Floquet mode corresponding to a quasienergy $\varepsilon_{\alpha}$.  
We note that, for integer $n$, 
\begin{align}
	\Phi_{\alpha'}(x,t) = \Phi_{\alpha}(x,t) e^{i n \Omega t } \equiv \Phi_{\alpha n}(x,t),
	\label{modesoln}
\end{align}
yields an identical solution to Eq.~\eqref{floquetsolution} with shifted quasienergy
\begin{align}
	\varepsilon_\alpha \to \varepsilon_{\alpha'} = \varepsilon_{\alpha} + n \hbar \Omega = \varepsilon_{\alpha n}.
	\label{quasishift}
\end{align}
Hence, the index $\alpha$ actually refers to a whole class of solutions indexed by $\alpha' = (\alpha,n)$ where $n = 0, \pm 1, \pm 2, \dots$  

The quasienergies therefore, are defined modulo $\hbar \Omega$ and can be mapped into a first Brillouin zone obeying $-\hbar \Omega/2 \leq \varepsilon < \hbar \Omega/2$. 
The quasienergy can be viewed as the time-periodic analogue to the quasi-momentum in Bloch's theorem of spatially periodic systems. 
The Floquet modes are eigenfunctions of the Floquet matrix $U$, which acts as a time evolution operator by stepping the solutions $\Psi_{\alpha}(x,t)$ forward in time by integer multiples of the driving period \cite{Reichl1992}

\begin{align}
	\Psi_{n}(x,t+T) = \sum_{m}U_{nm}(T) \Psi_{m}(x,t).
	\label{FloqMatrix}
\end{align}

In this work, we construct and diagonalise the Floquet matrix to compute the Floquet states of the one dimensional Schr\"odinger equation driven by representative external potentials.
While Floquet systems do not in general conserve energy, the \textit{time-averaged} energy
\begin{align}
	\bar E_{\alpha}&=\frac{1}{T}\int_{0}^{T}\Braket{\Phi_{\alpha}(t)|H|\Phi_{\alpha}(t)},
	\label{timeaven}
\end{align}
 is conserved and may be used to classify the states; for example the Floquet ``ground state'' is the state with the lowest time-averaged energy.  
\subsection{Numerical approach}
In all the cases discussed in this work, the numerical package XMDS \cite{XMDS} is used to simulate the  Schr\"odinger equation 
\begin{align}
i \hbar \p{}{t} \Psi(x,t) & = \left( -\frac{\hbar^2}{2m}\p{^2}{x^2} + V(x,t)\right) \Psi(x,t), \label{schro}
\end{align}
with $V(x,t)$ a time-periodic trapping potential.
We study a 1D system of length $L$ with periodic boundary conditions. We use a basis of plane waves and use a sufficient density of grid points to ensure numerical accuracy for the Floquet states of interest (we have typically used $L=16$ and $N=256$ lattice points for the cases considered here).

We consider potentials of the form $V(x,t) = V_0( x - c(t))$. 
By varying the form of $V_0(x)$ and the driving function $c(t)$, we are able to construct a range of time-averaged potentials.  
To find the Floquet states as a function of the driving period $T = 2\pi/\Omega$, we simulate the time evolution of a complete basis for a time interval of one period $T$.  From this we can construct the Floquet matrix $U$, which can then be diagonalised to find the Floquet states $\Phi_\alpha(x,t)$. 

For driving frequencies that are near resonance with energy spacings of the bare trapping potentials it is not possible to obtain numerically accurate results. 
In our simulations, there are regions in which the numerics clearly do not converge, and thus the Floquet states obtained by diagonalisation are not accurate.
By increasing the number of lattice points, it is possible to obtain more accurate Floquet states for a greater region of parameter space, but doing so is computationally expensive. 
In the Floquet states computed below, the regions of non-convergence are characterised by the states reaching the edge of the spatial grid and are correspondingly accompanied by a resonance spike in the time-averaged energy spectrum.

\section{One-dimensional ring potential} \label{sec:ring}
We first investigate the case of a  ring trap created by rapidly scanning a localised attractive potential in a circle.  This geometry is common in experiments  \cite{Lesanovsky2007,Henderson2009,Gildemeister2009,Gildemeister2010,Bell2016,Bell2018,Pandey2019} as it may be used for example in matterwave interferometry~\cite{Su2010}, sensitive gravimetry~\cite{Canuel2006}, rotation sensing~ \cite{Gustavson1997,Wu2007} and investigations of topological states of matter~\cite{Morizot2006,Yao2017}. In particular, we investigate a system similar to that studied by Bell~\emph{et al.}~\cite{Bell2016, Bell2018} who realised a ring trap for a BEC by circularly scanning an attractive optical dipole potential.  They performed a theoretical analysis of their system in order to understand unusual features in the atomic density in time-of-flight imaging, and found that it resulted from a non-trivial phase profile due to the time-averaged potential~\cite{Bell2018}.  We find that further insights are provided by applying a simple 1D Floquet analysis.

The trapping potential is \begin{align}
	V(x,t) & = V_{D}\exp\left(-\frac{(x-v t)^2}{2\sigma^2} \right),
	\label{ringtrap}
\end{align}
with $x-vt$ defined modulo $L$ such that we have periodic boundary conditions, and $v=L/T$ is  the speed of the potential such that it returns to its initial position after one driving period $T$. We consider an attractive potential of depth $V_D=-10$ and width $2\sigma^2 =1$ such that $\sigma \ll L$.
\begin{figure}[!t]
	\includegraphics[width=\linewidth,trim= {1.6cm 0.6cm 1.2cm 1.5cm},clip]{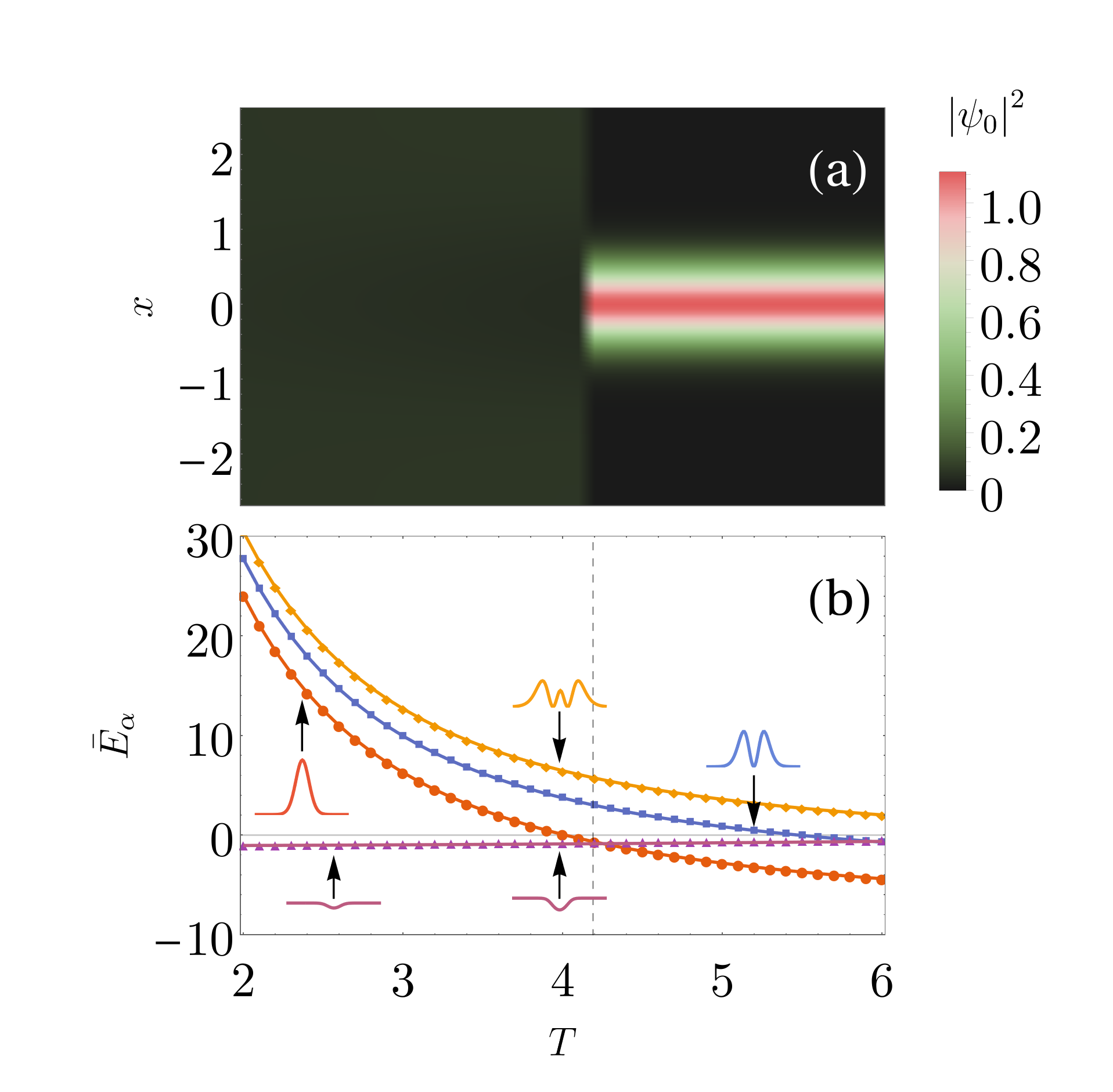}
	\flushright
	\caption{Localised to delocalised transition in 1D ring trap. {(a)} The Floquet ground state density  at $t=0$ for the 1D ring trap as a function of driving period. At $T=4.19$ the ground state  transitions from being homogeneous to the localised ground state of the attractive potential. 
	(b) The time-averaged energy spectrum [Eq.~\eqref{timeaven}] as a function of driving period.  
	Points are the results of Floquet simulations, and solid lines for the bound states are the result of the analytic theory given by Eq.~\eqref{totalen}.
	The solid line for the energy of the homogeneous state is from Eq.~\eqref{Madelung} with the phase given by Eq.~\eqref{phasepro}. 
	The black dashed line is the predicted transition point from equating Eqs.~\eqref{totalen} and \eqref{phasepro}. 
	Insets show the Floquet state density corresponding to each spectral curve. 	
	\label{Planestacked}}
\end{figure}

Since the moving potential spends the same amount of time at each point in space in one period, the time-averaged potential is simply a constant energy offset, independent of $x$
\begin{align}
	\bar{V}(x) &=\frac{\sqrt{2 \pi}V_D \sigma}{L},
	\label{timeavring}
\end{align}
and the Floquet states in the time-averaged limit $T \to 0$ are plane waves. Additionally, from Eq.~\eqref{ringtrap}, the states in the adiabatic limit $T\to \infty$ ($v\to 0$) will be the familiar eigenstates of a stationary Gaussian potential.
For our parameters, there are four bound states, and a number of unbound states with  winding numbers $w = 0, \pm 1, \pm 2$ etc.  The winding numbers are related to the topology of the ring and are a robustly conserved quantity. 
In accordance with our goal of understanding the transition from these Gaussian eigenstates to the plane-wave eigenstates in the time-averaged potential, we compute the Floquet states for a range of different driving periods, $T$.

\subsection{The transition between time-averaged and adiabatic limits}
In Fig.~\ref{Planestacked} we plot the Floquet ground state density as the scanning period $T$ is varied. As $T$ is increased, the Floquet ground state for the 1D ring system transitions discontinuously from an unbound, plane-wave like state, to the  localised ground state of the stationary potential. 

To learn more about how this transition manifests, in Fig.~\ref{Planestacked}(b) we show the the time-averaged energies [\eqreft{timeaven}]  and density profiles  for some representative Floquet states as a function of $T$: shown are the ground, first-excited and second-excited states in the adiabatic limit, and the ground state in the time-averaged limit. The energy of the bound states varies strongly with $T$, yet their density profiles are independent of $T$. In contrast, the energy of the unbound states is insensitive to the value of $T$ and is almost constant.  They also exhibit a density depletion at the instantaneous location of the attractive potential which deepens as $T$ increases. The size of this defect is slightly different for unbound states of different winding numbers. The sharp transition between the bound and unbound Floquet ground states occurs due to an exact crossing in the energy levels at $T = 4.19$. 

\subsection{Solutions in a Galilean boosted frame}
The trapping potential Eq.~(\ref{ringtrap}) is time-independent in the translating frame coordinates $x_c = x-v t$. 
This defines a Galilean boost, which transforms the Hamiltonian (see Appendix~\ref{appendix:Galileo}) 
\begin{align}
    H_c(x_c,t) &= \frac{p^2}{2m} - v p + \frac{1}{2} m v^2 + V(x_c),\\
    &= \frac{(p-mv)^2}{2m} + V(x_c).
\end{align}
Since the Hamiltonian $H_c$ is time-independent, it conserves energy and we can use standard separation of variables to find solutions of the form $\Psi(x_c,t) = \varphi(x_c)e^{-i E_c t}$
where $\varphi(x_c)$ obey the eigenvalue equation $H_c \varphi(x_c) = E_c \varphi(x_c)$
and $\varphi(x_c) = \varphi(x_\ell -v t)$ are Floquet modes: after a driving period  $\varphi(x_\ell -v T) = \varphi(x_\ell - L) = \varphi(x_\ell)$, since the spatial coordinate $x_\ell - vt$ is defined modulo $L$.
The energy in the lab frame $E$ is related to the translating frame energy $E_c$ by
\begin{align}
    E = E_c + v\Braket{p} - \frac{1}{2}mv^2.
    \label{totalen}
\end{align}

In the regime of fast scanning, the term containing the potential evolves significantly faster than the timescale over which the kinetic term evolves. 
We can therefore approximate the dynamics of the system by neglecting the kinetic energy term, and obtain an analytical approximation for the wave function. Although here we consider the non-interacting case of the linear Schrodinger equation, this approach can also be applied to nonlinear Schrodinger-type equations, for example the Gross-Pitaevskii equation for a weakly interacting Bose-Einstein condensate~\cite{Bell2018}, provided the external potential energy also dominates over the interaction energy.
We write the wave function in the Madelung form 
\begin{align}
	\Psi(x_c) = \sqrt{n(x_c)}e^{i \phi(x_c)}.
	\label{Madelung}
\end{align}
where the density $n(x_c) = |\psi(x_c)|^2$ and phase $\phi(x_c)$ are both real functions. 
Neglecting kinetic terms and inserting Eq.~\eqref{Madelung} into the Schr\"odinger equation (see Appendix \ref{appendix:madelung}) gives   
\begin{align}
 \partial_x \phi(x_c) =  \frac{V(x_c) - E'}{\hbar v}, 
	\label{phaseevo}
\end{align}
an ordinary differential equation for the phase, which can be readily solved to obtain
\begin{align}
	\phi(x_c) = \frac{V_D \sqrt{\pi} \sigma}{\sqrt{2} \hbar v} \left(\text{erf}(x_c/\sqrt{2}\sigma) - \frac{2}{L}x_c  \right)- \frac{2 \pi w x_c}{L},
	\label{phasepro}
\end{align}
where $w$ is the winding number~\cite{Bell2018}. The constant 
\begin{align}
    E' &= \frac{\sqrt{2\pi}V_D \sigma}{L} + v\frac{2\pi w  \hbar}{L}, \nonumber \\
    & = \bar{V} + v \braket{p}
\end{align}
approximates the energy of the non-kinetic terms in the boosted frame. 
A wave function with constant density and a phase profile given by \eqreft{phasepro} has lab frame energy
\begin{align}
    E(T) & = \bar{V} + \frac{(2 \pi)^2 \hbar ^2 w^2 }{2mL^2}+ \frac{ \sigma  T^2 V_D^2}{m L^3}\left(\frac{\sqrt{\pi }}{2}-\frac{\pi  \sigma }{L}\right), \label{eq:Eofplane}
\end{align}
i.e., it has a constant offset $\bar V$, the kinetic energy of a plane-wave with $p = 2\pi \hbar w/L$, and a term which grows as $T^2$ which represents the contribution of the phase profile to the energy. 
We plot \eqreft{eq:Eofplane} as solid line in Fig.~\ref{Planestacked}(b).
The ground state solution has winding number zero, as this minimises the energy. 

We are now in a position to understand the behaviour in Fig.~\ref{Planestacked}. 
At $T\to \infty$ ($v=0$), we are in the adiabatic limit, and the lab frame energy is the same as the stationary problem, i.e., $E=E_c$.
The bound states, which are real and non-degenerate, must have $\Braket{p_c}=0$ in the centre of mass frame. That is, they have $\braket{p}- mv = 0$. 
Thus, from \eqreft{totalen}, we see that overall, their energy changes from the stationary problem simply by the addition of the kinetic term  $\frac{1}{2}mv^2$. The bound states are sensitive to the driving period $T$ in the lab frame, and insensitive in the translating frame. 

In contrast, from \eqreft{eq:Eofplane} it can be seen that the unbound states are quite insensitive to the drive in the lab frame. 
As $T \to \infty$, the bound states approach their translating frame energy $E_c$, whereas the unbound states \textit{increase} their energy slightly due to a combination of an increasing variance $\sigma_p$ as well as a growing depletion in their density profile.
Hence, we have a crossover in the spectrum.

\subsection{Quantification of time-averaging}
\begin{figure}[!t]
    \centering
    \includegraphics[width=\columnwidth]{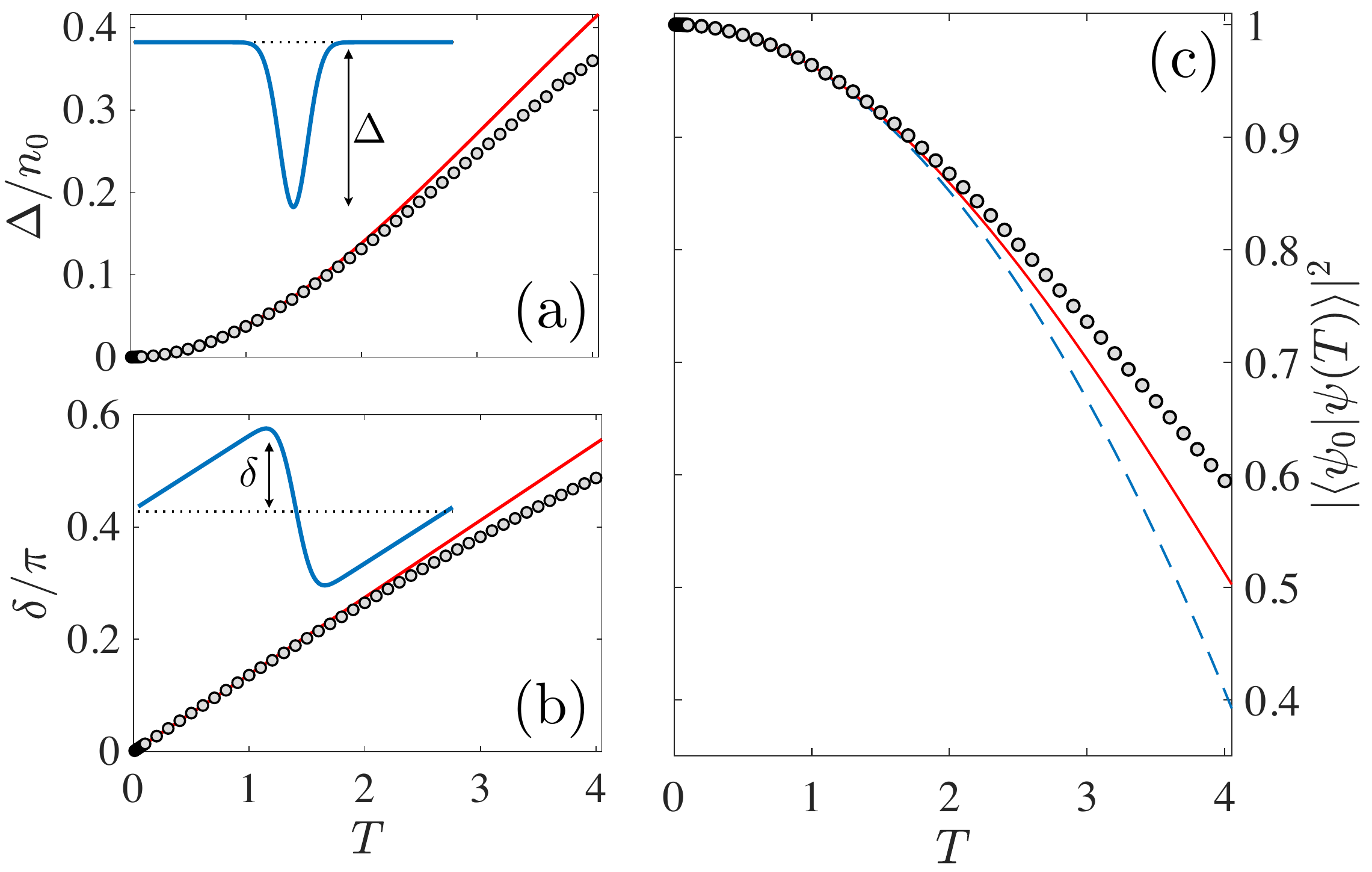}
    \caption{Properties of the Floquet ground state vs. scanning period $T$ in the time-averaged limit $T\rightarrow 0$. (a) Depth of the density defect $\Delta$. (b) Height of the phase defect $\delta$. (c) Fidelity of the Floquet ground state with the $k=0$ plane-wave ground state of the time-averaged limit. Circles show numerical results. Solid lines show the predictions given by Eq.~\eqref{Madelung}. The blue dashed line in (c) shows the analytical approximation Eq.~\eqref{eqn:phasefidelity}.
}
    \label{planefidel}
\end{figure}

For values of $T<4.19$ the Floquet ground state is close to being homogeneous, but exhibits small deviations in the form of a density defect and a nonlinear phase profile that become larger as $T$ increases. A natural question that remains is how well these states approximate the homogenous ground state in the time-averaged limit.  By comparing the Floquet ground state $\psi(x,T)$ at some driving period $T$ to the ground state of the time-averaged potential, $\psi_0(x) = L^{-1/2}$, we can quantify the quality of the time-averaged approximation for a given  period $T$ through the fidelity $f = \Braket{\psi_0 | \psi(T)}$.  
We can obtain an analytic approximation for the phase step height of the ground state as a function of driving period
\begin{align}
	\delta(T) = \frac{V_D \sqrt{\pi} \sigma}{\sqrt{2} \hbar L} \left(\text{erf}(x_M/\sqrt{2}\sigma) - \frac{2}{L}x_M  \right) T,
	\label{phasestepanalytic}
\end{align}
where $x_M = -\sqrt{2} \sigma  \sqrt{\log \left(L/\sqrt{2 \pi } \sigma \right)}$ is the $x$ coordinate where the phase reaches its maximum value. In the inset of Fig.~\ref{planefidel}(b), we show the phase profile, along with the definition of the phase step height $\delta = \max(\phi)$.

In the boosted coordinates, the continuity equation (see Appendix \ref{appendix:madelung}) takes the form
\begin{align}
    -v \, \partial_x n + \partial_x (n u) = 0, 
\end{align}
where $u = \hbar \partial_x \phi/m$.  
This is an ordinary differential equation for the density $n(x)$ and can be solved to obtain 
\begin{align}
    n(x_c) &= \frac{A}{u(x_c)-v},
\end{align}
where the integration constant
$A = 2 \pi  w \hbar /mL^2 -T^{-1}$ is determined since the density must be normalised to unity. 
We can then obtain an expression for the depth of the density defect $\Delta/n_0$
\begin{align}
    \frac{\Delta}{n_0} = \frac{L T^2 V_D}{T^2 V_D \left(L-\sqrt{2 \pi } \sigma \right)-L \hbar  \left(L^2+2 \pi  T w\right)}.
\end{align}
Using the full wave function $\Psi = \sqrt{n}e^{i \phi}$, we can compute the ground state fidelity as a function of $T$, which is shown as a solid red line in Fig. \ref{planefidel}(c).
By taking a series expansion to second order, we can obtain an analytic approximation for the fidelity with the time-averaged ground state
\begin{align}
    f &=1 - \frac{\pi  \sigma ^2 T^2 V_D^2}{L^2 \hbar ^2} \left(\frac{\sigma ^2}{L^2}-\frac{\sigma }{\sqrt{\pi } L}+\frac{1}{12} \right),
    \label{eqn:phasefidelity}
\end{align}
which is plotted as a dashed blue line in Fig. \ref{planefidel}(c).

The dip in the density and spatially varying phase profile will affect the dynamics of atoms in experimental time-averaged traps. 
From our 1D Floquet analysis, we are able to qualitatively reproduce the features found experimentally by Bell \textit{et al.} who made use of a 2D Gross-Pitaevskii simulation to investigate the phase profile a BEC in a time-averaged ring trap~\cite{Bell2016, Bell2018}.   
Specifically, we find the same homogeneous Floquet states in the time-averaged limit as well as the phase profile identified as responsible for the ``kink'' in the ring produced in the experiment.

\section{Sinusoidally driven trapping potentials}
\label{sec:nonGalilean}
While the 1D ring of Sec.~\ref{sec:ring} can be understood using Floquet theory, it was not  necessary as there was a Galilean boost which rendered the problem time-independent.  Hence we could use exact diagonalisation to understand the different energy dependence on $T$ for the bound and unbound states, leading to a level crossing in the energy spectrum. 
However, this is only one way in which the localised to delocalised transition can occur. 
In this section we investigate systems for which there is no time-independent frame of reference, and so a full Floquet analysis is required. We illustrate the different ways that the Floquet states transition from the adiabatic to the time-averaged limits with three representative trapping potentials.

\subsection{Driven harmonic oscillator }
\label{sec:harmonic}
We first consider perhaps the simplest theoretical model which breaks Galilean invariance and has ground states which are bound in both the time-averaged and adiabatic limits. 
\begin{figure}[!t]
	\includegraphics[width=\linewidth,width=\linewidth,trim= {2.7cm 0.5cm 2.7cm 0.5cm},clip]{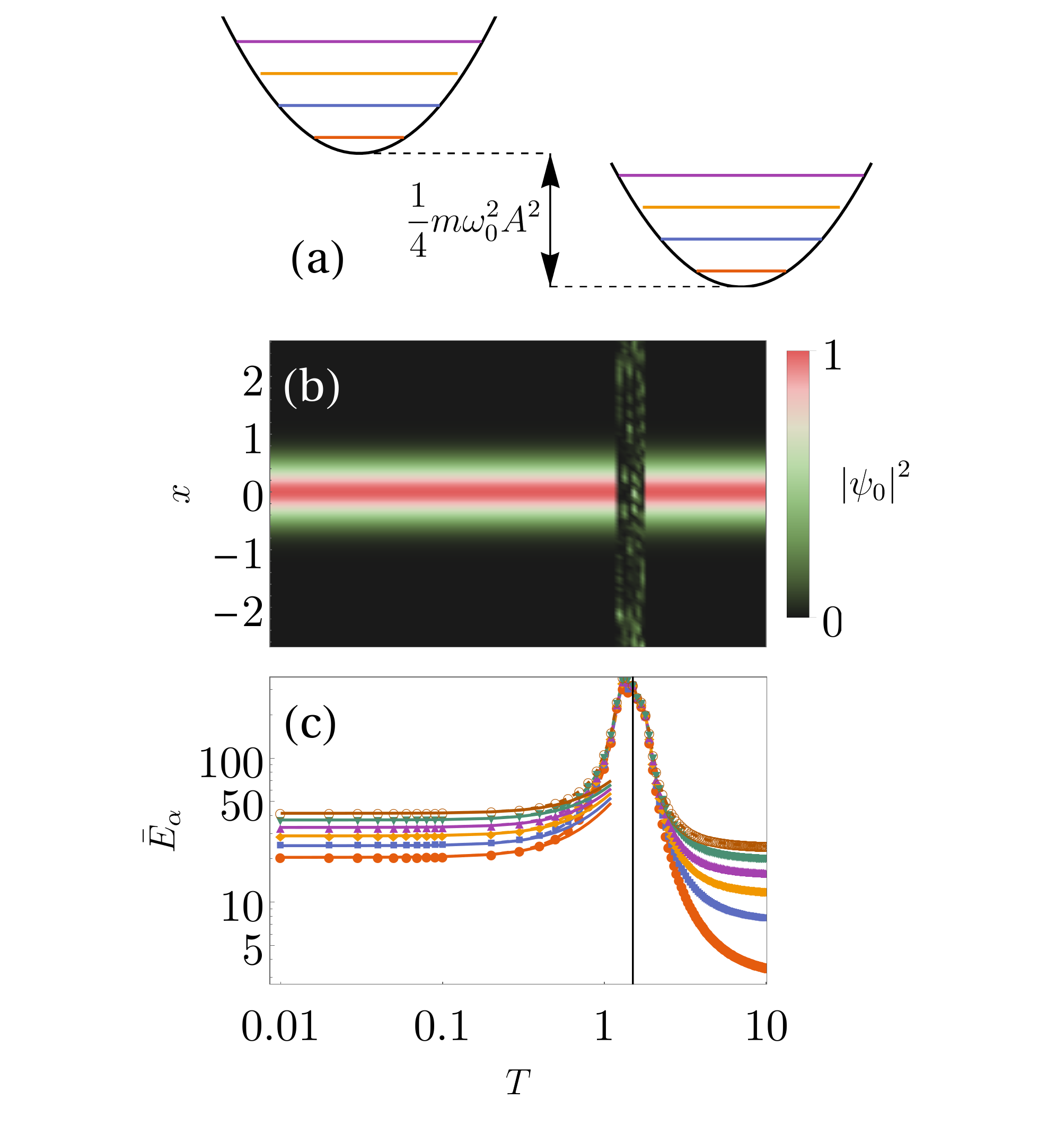}
	\caption{
 Floquet analysis of the driven harmonic oscillator. 
		(a) A schematic of the time-averaged (left) and adiabatic (right) driven harmonic oscillator, with the lowest four energies indicated.
	(b) The Floquet ground state density for the driven harmonic oscillator with $\omega_0 =2\pi/ 1.5$.
	In a small region around $T_\text{res}=1.5$ the strong resonance means that the numerical results for the density have not converged. 
	Outside of this range, the numerics are stable and we obtain the expected harmonic oscillator ground state density.
	(c) The time-averaged energy spectrum as a function of driving period for the same parameters (markers).  	The solid lines are the result of an inverse frequency expansion to second order in $T$.
	The black vertical line at $T=1.5$ indicates the location of the resonance. 
	\label{HarmonicSpectrum}}
\end{figure}
We sinusoidally drive the position of a 1D harmonic oscillator 
\begin{align}
	V(x,t) = \frac{1}{2}m\omega_0^2\left( x + A \sin(\Omega t)\right)^2,
	\label{HarmSin}
\end{align}
where $\Omega = 2 \pi/T$ is the frequency of the drive.  We choose the amplitude $A=2$ and harmonic oscillator frequency $\omega_0 = 2 \pi/1.5$  (i.e the harmonic oscillator period is $T=1.5$).
This potential has time-average
\begin{align}
	\bar{V}(x) 	&=\frac{1}{2}m\omega_0^2 x^2 + \frac{1}{4}m\omega_0^2 A^2,
\end{align}
i.e, it is simply the same harmonic oscillator shifted up in energy by the constant $\frac{1}{4} m \omega_0^2 A^2$ as can be seen in Fig \ref{HarmonicSpectrum}(a).
This system will allow us to explore some additional features of Floquet systems that are important to the transition between the adiabatic and time-averaged limits. 

\begin{figure}[!t]
	\includegraphics[width=\columnwidth]{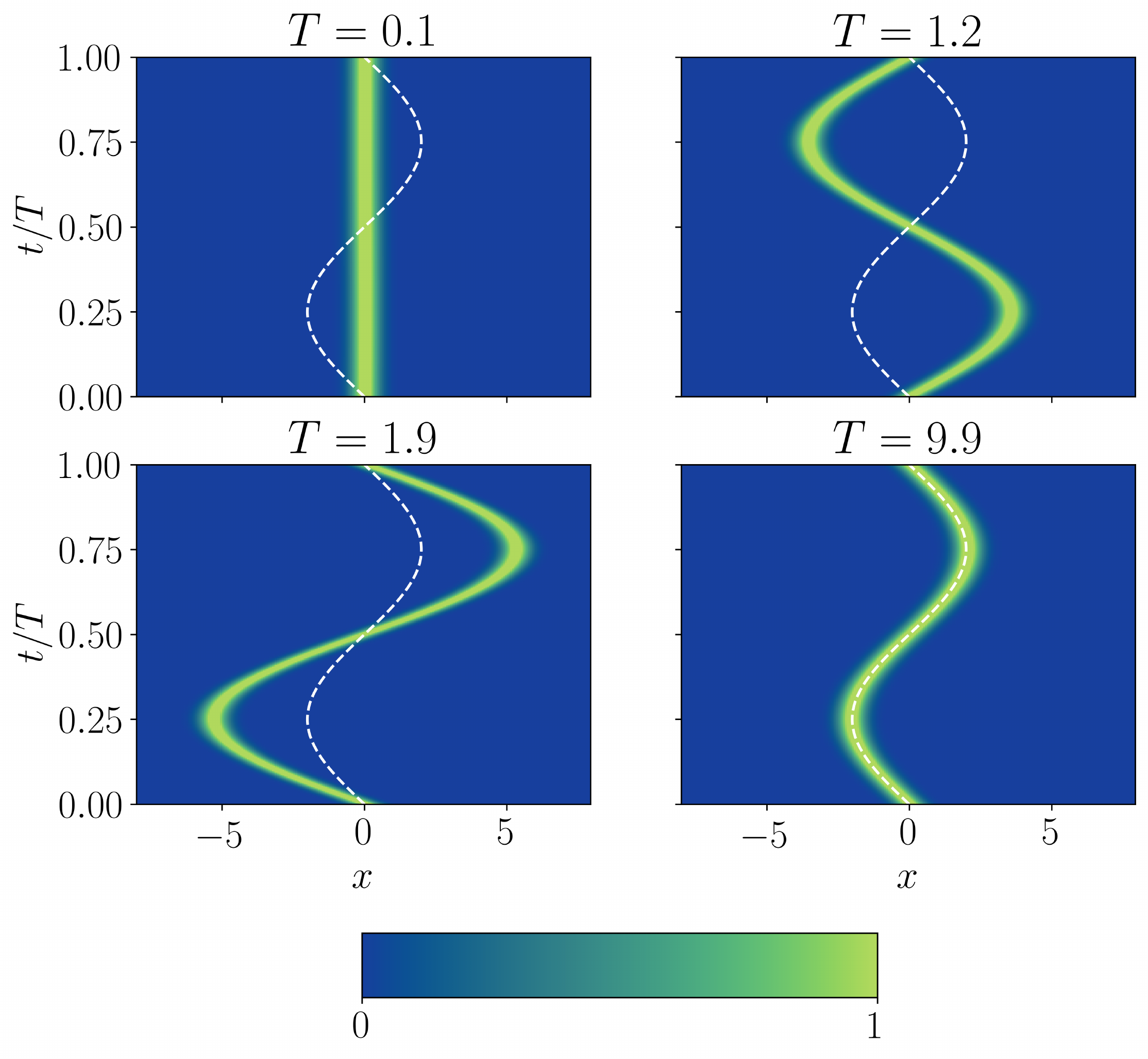}
	\caption{Illustration of the relative motion of the Floquet ground state density and centre of the trapping potential as the driving period T is increased.
		(a) $T=0.1$, near the time-averaged limit. 
		(b) $T=1.2$, just before resonance, the Floquet state undergoes large amplitude centre-of-mass oscillations and is out of phase with the trapping potential. 
	(c) $T=1.9$, after the resonance, the Floquet state  exhibits large amplitude centre-of-mass oscillations and is in phase with the trapping potential
	(d) $T=9.9$, where the Floquet state is far from resonance and follows the motion of the trapping potential. 
	\label{harmphaselock}}
\end{figure}

The results of a Floquet analysis of this potential are summarised in Fig.~\ref{HarmonicSpectrum}.
In Fig.~\ref{HarmonicSpectrum}(b) we can see that the Floquet ground state takes the form of a harmonic oscillator ground state in both the $T\rightarrow0$ and $T\rightarrow\infty$ limits.   Since the energy level spacings for the harmonic oscillator are constant, a collective resonance occurs in the region of $T_\text{res}=1.5$, which is clearly seen in the time-averaged energy spectrum in Fig.~\ref{HarmonicSpectrum}(c).
In a narrow region around the resonance point, the energies are too high for the states to be accurately calculated by our simulation, which results in the noisy region near $T=1.5$ in Fig.~\ref{HarmonicSpectrum}(b).

In Fig.~\ref{HarmonicSpectrum}(c), we have also plotted the results of an inverse frequency expansion \cite{Rahav2003,Goldman2014, Eckardt2015,Itin2015} up to second order in $T$, which approximates the energy spectrum perturbatively in powers of $1/\Omega$. (For more details see Appendix~\ref{appendix:invfreq}.) We find that the energy is well approximated with quadratic growth as $T$ approaches $T_\text{res}$. 

For this potential the collective resonance marks the localised to delocalised crossover.  In Fig~\ref{harmphaselock}, for $T<T_\text{res}$ we can see that the Floquet ground states oscillate out of phase with the centre of the potential, (white dashed line), while for $T>T_\text{res}$, they are in phase with it. As would be expected, the effects of the driving are more significant for $T$ close to $T_\text{res}$.
For $T=0.1$, we are clearly in the time-averaged limit, as there is almost no centre-of-mass oscillation, and the fidelity (see Fig.~\ref{fig:combinedFidelity}) is close to unity. In the opposite limit for $T=9.9$ the system is close to the the adiabatic limit, and the Floquet ground state follows the oscillating potential..

In contrast to the ring potential of Sec.~\ref{sec:ring}, the Floquet states of the driven harmonic oscillator are the same in the two limits, so there are no energy level crossings as $T$ is varied.

\subsection{Driven linear potential}
\label{sec:abs}
In the case of the driven harmonic oscillator, the Floquet states in the time-averaged and slow-moving limit were the same. 
This allowed us to highlight the role that the collective resonance played in the transition between the two regimes of interest. 
Here we consider the potential 
\begin{align}
	V(x,t) & = V_D\left| x + A \sin\left( \frac{2\pi t}{T} \right) \right|,
\end{align}
which is harmonic in the time-averaged limit, but has the form $V(x) \sim |x|$ in the slow-moving limit.  We choose the numerical parameters $V_D=10$ and $A=3$. 
The analytic expression for the time-averaged potential is
\begin{align}
	\frac{\bar{V}(x)}{V_D}& =  
	\begin{array}{cc}
		 \Bigg\{ & 
			 \begin{array}{cc}
				   2 \left[\sqrt{A ^2-x^2}+x \sin ^{-1}\left(x/A\right)\right]/\pi,  & \,|x| \leq A, \\
				  \,\left| x\right|,  &|x| > A. \\ 
			   \end{array}
			    \\
		    \end{array}
\end{align}
In the time-averaged limit the Floquet states are well-approximated by  harmonic oscillator eigenstates.
Performing a series expansion in $x$ gives 
\begin{align}
	\frac{\bar{V}(x)}{V_D} \approx \frac{2 A}{\pi} + \left(\frac{1}{\pi A} \right)x^2 + \mathcal{O}(x^4)
\end{align}
which yields a harmonic oscillator frequency of $\omega_{HO} = (2 V_D/\pi A)^{1/2}$. 

\begin{figure}[!t]
	\includegraphics[width=\linewidth,trim= {3.1cm 0 2.5cm 0},clip]{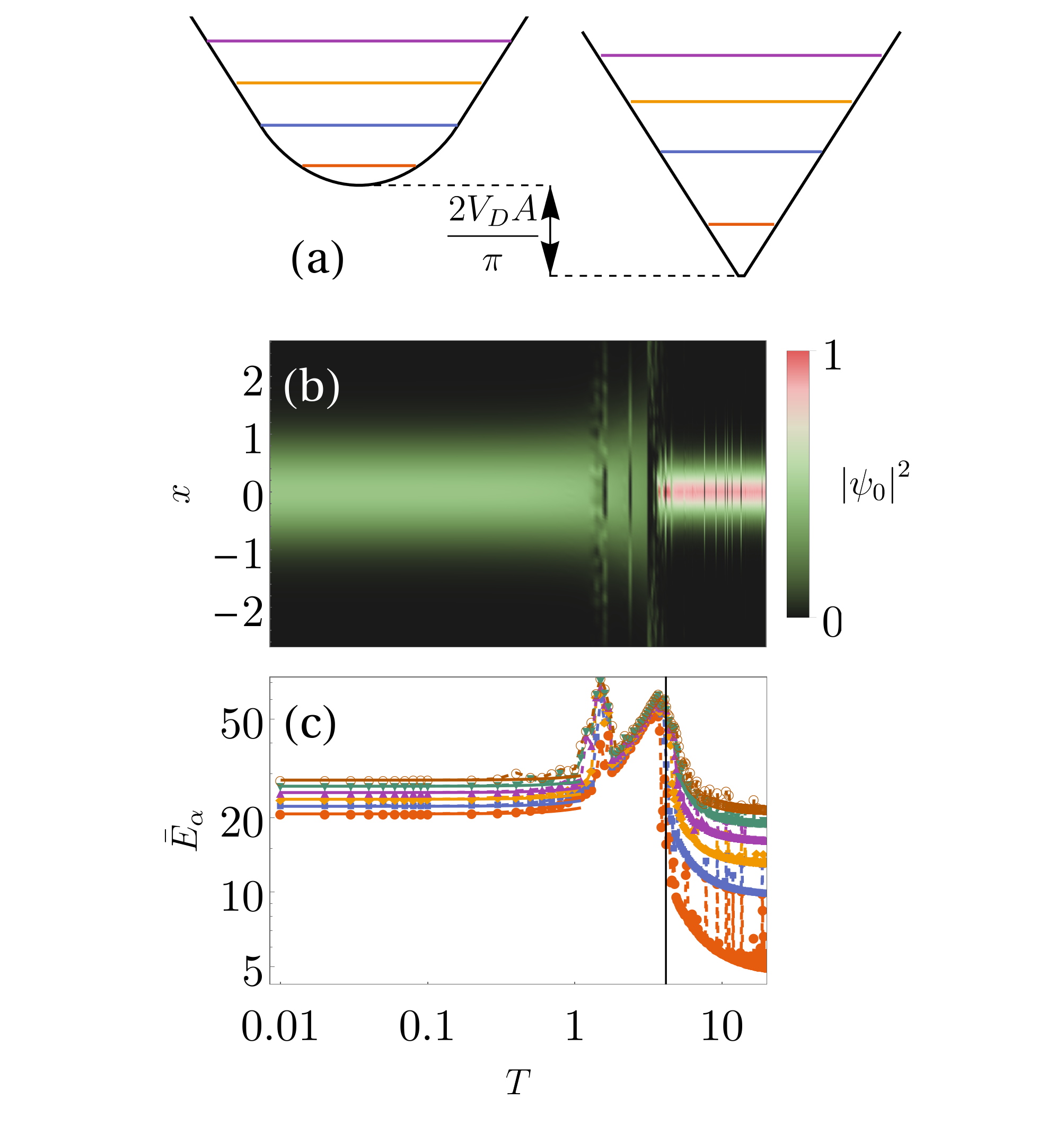}
	\caption{Large collective resonance and emergence of ``photon'' resonances for the driven $|x|$ potential. 
	(a) A schematic of the time-averaged (left) and adiabatic (right) driven $|x|$ potential, with the lowest four energies indicated.
		(b)
		The Floquet ground state density for the driven $|x|$ potential as a function of driving period $T$.
	(c)
	The time-averaged energy spectrum for the driven $|x|$ potential as a function of driving period $T$.
	The states undergo a collective resonance, the position of which can be predicted by the energy spacings for low $T$ (vertical black line). 
	The numerics do not converge in small regions around the collective and photon resonance points where the Floquet state reaches the edge of the $x$ grid. 
		Solid lines are the result of an inverse frequency expansion, where the third order term has been fitted to the Floquet simulation. 
		\label{absEnergy}}
\end{figure}

In Fig.~\ref{absEnergy}(b) we plot the Floquet ground state density as a function of the period $T$, and again observe the presence of a collective resonance peak due to the equal energy spacing of the time-averaged harmonic oscillator potential. 
As in the harmonic oscillator case (Sec.~\ref{sec:harmonic}), for $T \ll T_{\mathrm{res}}\approx 2\pi/\omega_{HO}$ the Floquet ground state is delocalised and oscillates out of phase with the potential, and vice versa for $T\gg T_{\mathrm{res}}$.  The collective resonance is again the most significant contribution  to the decrease in fidelity with increasing $T$ and thus the transition to the adiabatic limit. 
A new feature for this potential, however, is the presence of multi-``photon'' resonances ~\cite{Eckardt2015}, which appear as smaller resonances outside of the main resonance peak. 
These resonances are due to avoided crossings in the \textit{quasienergy} spectrum as a result of hybridisation of states in one ``photon'' block with another. 
In terms of the extended Hilbert space $\mathcal{H} \otimes \mathcal{T}$, an $N$-``photon'' resonance results from the coupling of two Fourier modes $k, \ell$ with $k - \ell = N$ in the space $\mathcal{T}$.
These resonances are not captured to any order by inverse frequency expansion, which explicitly removes the matrix elements responsible for the coupling of one ``photon'' block to another, dealing only with diagonal elements in the extended space. 
These resonances are important for experiments, as they result in a breakdown of adiabatic following \cite{Hone1997} of the Floquet states and physically represent heating due to energy transferred from the drive \cite{Eckardt2015}.
Thus, experiments should choose parameters which avoid this adiabatic breakdown, both for preparing and measuring time-periodic systems.   

\subsection{Driven quartic double well}
\label{sec:quartic}
The previous two trapping potentials considered in this section have had harmonic oscillator eigenfunctions in the time-averaged limit. 
This led to a collective resonance occurring and a clear separation between localised and delocalised regimes. 
Here we consider a trapping potential which still admits bound states in both the time-averaged and slow-moving limits, but has an anharmonic spectrum everywhere. 
\begin{figure}[!t]
	\includegraphics[width=\linewidth,trim= {2.6cm 0 2.6cm 0},clip]{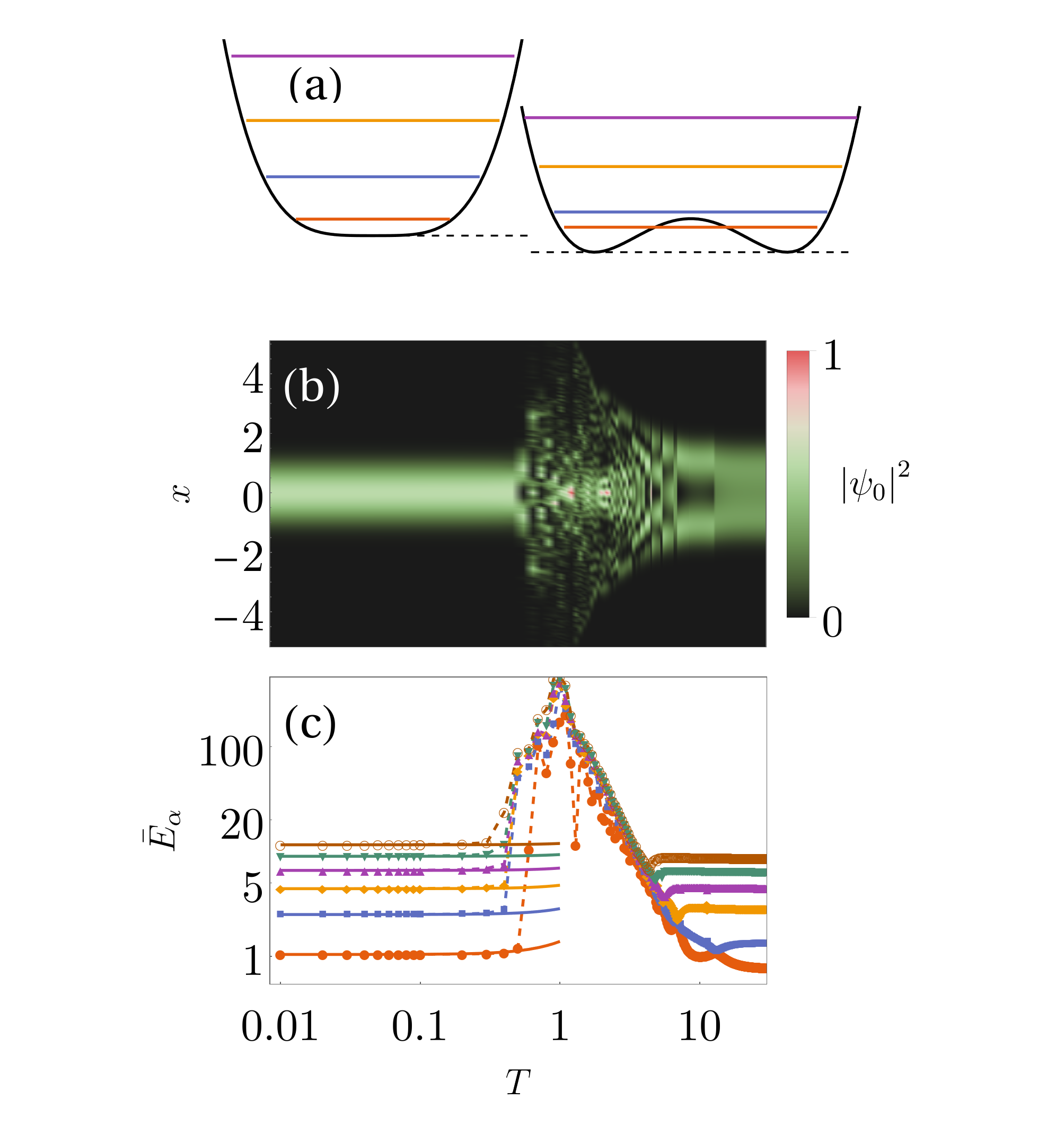}
	\caption{The quartic double well displays a quasi-collective resonance as well as photon resonances.
	(a) The time-averaged (left) and adiabatic (right) potentials.
		(b) 
		The density of the Floquet ground state for the driven double well potential as a function of $T$.
		At large $T$ we recover the two-peaked ground state of the double well potential. 
(c) The time-averaged energy spectrum for the driven double well potential as a function of $T$. 
The numerics do not converge in regions where the Floquet state reaches the edge of the $x$ grid. 
Solid lines are the result of an inverse frequency expansion to third order.}
	\label{QuarticStacked} 
\end{figure}

We consider the quartic double well potential and subject it to sinusoidal driving
\begin{align}
	V(x',t) &= A\left( e_0 x'^4 -x'^2 \right),
	\label{eq:quartic}
\end{align}
where the coordinate $x' = x+ \epsilon \sin(\Omega t)$, and we choose the numerical values $A=1.5$ and $e_0=0.36$, so that the potential has a double-well shape in the adiabatic limit. 
This potential Eq.~(\ref{eq:quartic}) has time-average
\begin{align}
	&\bar{V}(x) 
&= A \left( e_0 x^4 + \frac{1}{8}\epsilon^2\left( 3e_0 \epsilon^2 -4 \right)+ x^2 \left( 3 e_0 \epsilon^2-1 \right) \right).
\label{quarticavg}
\end{align}

The results for this potential are summarised in Fig.~\ref{QuarticStacked}, where it can be seen that this system transitions in a qualitatively different manner to the previous two.  Since there is no collective resonance, there is no single point after which the states become localised in the potential.
The Floquet ground state density as a function of the driving period $T$ is shown in  Fig.~\ref{QuarticStacked}(b).  A number of different ground states densities are apparent, and there are regions of collective resonances, photon resonances and mixing between the states of each limit.  
Since the energy level spacings of the lowest energy states are reasonably similar in magnitude, there is a quasi-collective resonance where many of the states are destroyed at similar values of $T$.
As before, it is possible to compute an approximation to the high frequency energy spectrum via an inverse frequency expansion, and we find good agreement at second order [solid lines in Fig.~\ref{QuarticStacked}(c)]. 

\subsection{Behaviour of fidelity}
For the harmonic systems considered in Secs.~\ref{sec:harmonic} and \ref{sec:abs}, which have equal energy level spacings in the time-averaged limit, a collective resonance marks the localised to delocalised transition point. 
For anharmonic systems as in Sec.~\ref{sec:quartic}, this collective resonance does not occur and as such, there is no clear localised to delocalised transition point and states are able to undergo mixing in intermediate regions of parameter space.  

In Secs \ref{sec:abs} and \ref{sec:quartic}, we highlighted the impact of so-called ``photon'' resonances on the transition and general dynamics of Floquet systems. 
For experiments, regions of photon resonance should be avoided, as it leads to uncontrollable heating which will ultimately destroy the system. 

\begin{figure}[!t]
	\includegraphics[width=\linewidth,]{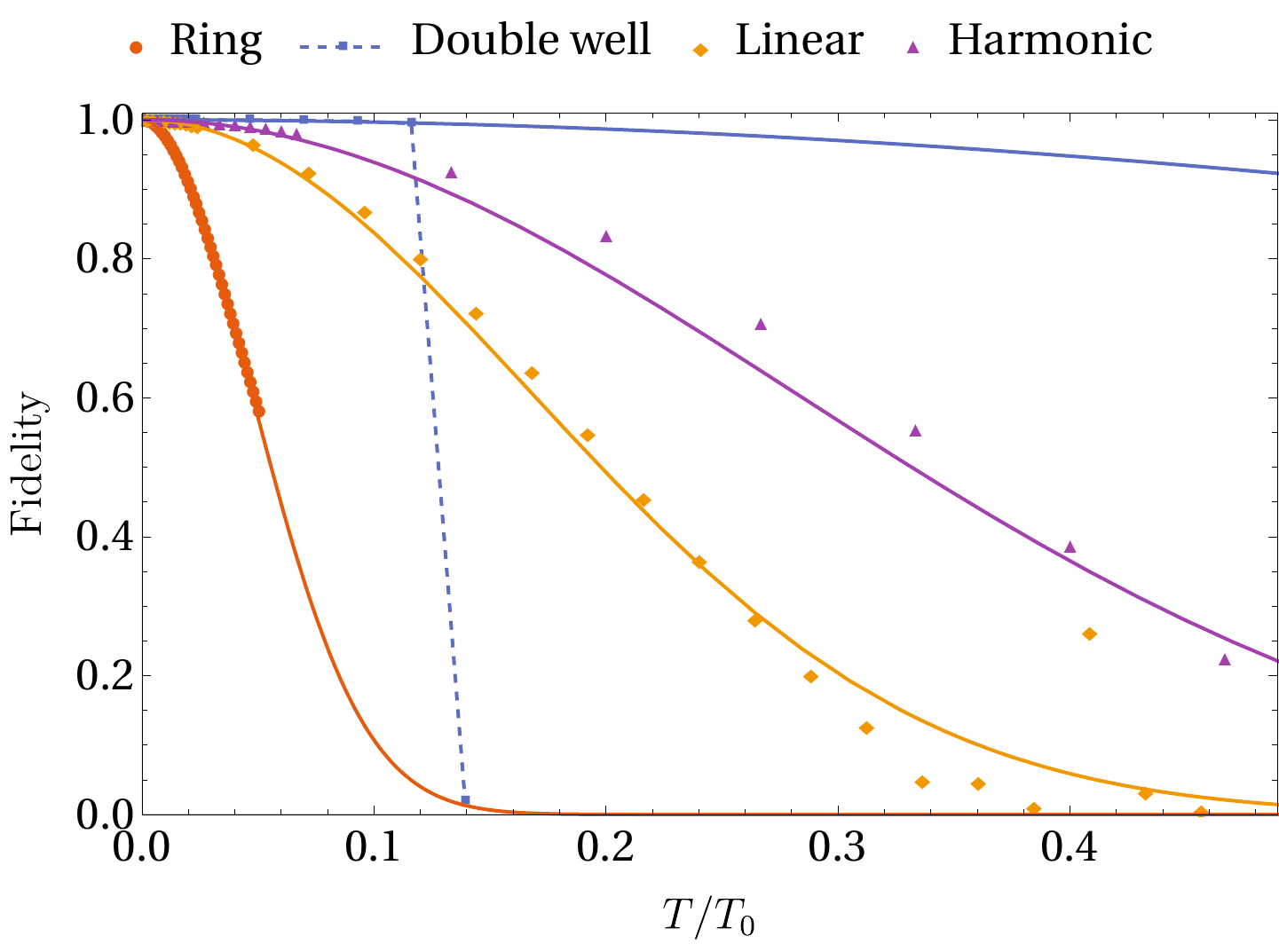}
	\flushright
	\caption{Comparison of the fidelity of the Floquet ground state with the time-averaged limit for as a function of scaled driving period for the three potentials considered in this section.  Solid lines are a Gaussian fit to the data. Dashed lines have been added between the quartic double well data points to aid the eye.
	\label{fig:combinedFidelity}}
\end{figure}

The fidelity between the time-averaged ground state provides a quantitative measure of the degree to which a Floquet state is in the time-averaged limit. 
In Fig.~\ref{fig:combinedFidelity} we plot the fidelity as a function of scaled driving period $T/T_0$, where 
the characteristic driving period $T_0 = h/\Delta E$ is determined by the energy difference $\Delta E$ between the ground and first excited states.
The fidelity decay for the Floquet systems studied is well approximated by a Gaussian (solid lines in Fig. \ref{fig:combinedFidelity}). For the quartic double well, the fidelity remains close to unity until a photon resonance sharply destroys the Floquet ground state.

\section{Conclusions}
\label{sec:conclusions}
We have studied time-averaged potentials for ultracold atoms using the method of Floquet analysis. 
We first considered a 1D ring potential formed by an attractive Gaussian beam scanned at a constant velocity. 
Due to the Galilean invariance of this potential, it was possible to obtain Floquet states for this system by transforming to a time-independent frame of reference. 
We have built on the earlier work of Bell \emph{et al.}~\cite{Bell2018} to demonstrate how the Floquet states of this system change as a function of the driving period, and have derived several approximate analytic results.  Our results provide further insights into the nature of this time-averaged potential.

We then performed a Floquet analysis of three other 1D potentials where the position of the trap was driven sinusoidally. 
These examples illustrated the effects of a collective resonance when harmonic oscillator potentials were driven near their resonant frequency, and photon-resonance effects due to coupling between individual Floquet states.  We performed an analytic inverse-frequency expansion in the fast-moving limit that agreed well with our numerical results.
We used the fidelity to provide a quantitative measure of the degree to which these systems approximated the time-averaged limit.

Our results clearly illustrate the transition of the ground state density from  adiabatic following of a moving potential to being delocalised in the time-averaged potential. They show the resonances that can lead to heating that when making use of of driven trapping potentials for ultracold atoms.  

The fidelity between the time-averaged state and the Floquet states decays proportional to $T^2$ for small $T$ and is approximately Gaussian at larger $T$. This fidelity decay is a common feature of all the systems studied here and provides a more quantitative means for determining an appropriate scanning rate than the rule-of-thumb $\Omega \gg \omega$.

In this work, we restricted ourselves to potentials which are monochromatically driven in time, which have a single frequency of driving. For non-monochromatic driving, the presence of multiple driving frequencies drastically increases the complexity of the resulting dynamics, and as such is an important consideration for future work.  

\acknowledgments
This research was supported by the Australian Research Council
  Centre of Excellence in Future Low-Energy Electronics Technologies
  (project number CE170100039) and funded by the Australian
  Government.  

\appendix
\section{Galilean transformation to a time-independent Hamiltonian} 
\label{appendix:Galileo}
The Schr\"{o}dinger equation is
\begin{align}
        i \hbar \p{ \Psi_\ell(x_\ell,t)}{t} = \underbrace{\left[ -\frac{\hbar}{2m} \p{^2}{x_\ell^2}+V(x_\ell,t) \right]}_{H(x_\ell,t)}\Psi_\ell(x_\ell,t),
        \label{SchroLab}
\end{align}
where $\Psi_\ell(x_\ell,t)$ is the wave function in the laboratory frame, and $x_\ell$ is the laboratory frame position coordinate.

The coordinate transformation to the moving frame
\begin{align}
        x_c = x_\ell - v t,
\end{align}
transforms the various functions according to
\begin{align}
        \Psi_{\ell}(x_\ell) & = \Psi_\ell(x_c + vt),\\
        \p{ }{t} \Psi_\ell(x_\ell, t ) &= \p{ }{t} \Psi_\ell(x_c + vt,t) - v \p{ }{x_c} \Psi_\ell(x_c+vt,t), \\
        \p{ }{x_\ell} \Psi_\ell(x_\ell,t) &= \p{ }{x_c} \Psi_\ell(x_c+vt,t),
        \label{}
\end{align}
so that Eq.~\eqref{SchroLab} becomes
\begin{align}
        i \hbar \p{}{t}\Psi_\ell = \underbrace{\left[ -\frac{\hbar^2}{2m}\p{^2}{x_c^2} + i \hbar v \p{ }{x_c} + V(x_c+vt,t) \right]}_{H'(x_c,t)}\Psi_\ell.
        \label{SchroComov}
\end{align}
Now recall that $\p{}{x_c} = i p$, where $p = \hbar k$ is the conjugate momentum, and we can write
\begin{align}
        i \hbar \p{}{t}\Psi_\ell = \left[ \frac{p^2}{2m} - v p + V(x_c + vt, t) \right]\Psi_\ell .
\end{align}
Since the potential is of the form $V(x,t) = V(x-vt)$, the Hamiltonian becomes time-independent and can be solved via separation of variables, giving the eigenvalue equation
\begin{align}
        \left[ \frac{p_c^2}{2m} - v p_c + V(x_c) \right]\varphi(x_c) = E' \varphi(x_c).
        \label{eq:boostEigvalEq}
\end{align}
The bound state solutions of this eigenvalue problem (which move with the potential in the laboratory frame) have $\braket{p} = mv$. In the moving frame, they have $\braket{p_c} = 0$, i.e $p_c = p-m v$.
So the Hamiltonian in the moving frame (in terms of the lab momenta) is  
\begin{align}
        H_c = \frac{(p - mv)^2}{2m} + V(x_c),
        \label{H_COM}
\end{align}
which is simply $H' + \frac{1}{2}mv^2$. 
Since the addition of a constant does not change the dynamics, we can solve the moving frame problem given by Eq.~\eqref{H_COM}.

\section{Madelung transformation for a boosted Hamiltonian} \label{appendix:madelung}
Inserting the Madelung form $\varphi(x_c) = \sqrt{n(x_c)}e^{i \phi(x_c)}$ into \eqreft{eq:boostEigvalEq} and equating the imaginary components yields
\begin{align}
    -v \, \partial_x n(x) + \partial_x \left( n(x) u(x)\right)=0,
    \label{eq:continuity}
\end{align}
where $u(x) = \hbar/m \,  \partial_x \phi(x)$. 
Equating the real parts gives
\begin{align}
    V(x) - E' -m v u + \frac{1}{2}m u^2 -\frac{\hbar^2}{2m} \frac{\partial_x^2 \sqrt{n}}{\sqrt{n}}=0,
\end{align}
where $E'$ is a constant independent of $x$.
We assume $V \gg \frac{1}{2}m u^2 ,\frac{\hbar^2}{2m} \frac{\partial_x^2  \sqrt{n}}{\sqrt{n(x}} $ to obtain
\begin{align}
    \partial_x \phi(x) =  \frac{V(x) - E'}{\hbar v}.
\end{align}
\section{Inverse frequency expansion}
\label{appendix:invfreq}
Here we outline the calculation for the inverse frequency expansion for the driven harmonic oscillator. 
We follow Ref.~\cite{Goldman2014}, although we note that the same result may be obtained with equivalent methods described in Refs.~\cite{Rahav2003, Itin2015, Eckardt2015}. 

The effective Hamiltonian given in \cite{Goldman2014} is
\begin{align}
	H_{\text{eff}}^{(0)} &= H_0, \\
	H_{\text{eff}}^{(1)} & = \frac{1}{\Omega} \sum_{j=1} \frac{1}{j}\left[ V^{(j)},V^{(-j)} \right],\\
	H_{\text{eff}}^{(2)} &= \frac{1}{2\Omega^2}\sum_{j=1}^{\infty} \Big( \left[ \left[ V^{(j)},H_{0} \right],V^{(-j)} \right] \nonumber \\
	& + \left[ \left[ V^{(-j)},H_{0} \right],V^{(j)} \right]  \Big),
\end{align}
where $H_0$ is the time-averaged Hamiltonian, and $V^{(j)}$ are the Fourier components of the driving potential. 
Recall the Hamiltonian for the driven harmonic oscillator is 
\begin{align}
	 H(x,t) = \frac{p^2}{2m} + \frac{1}{2}m\omega_0^2 \left( x+A \sin(\Omega t) \right)^{2},
\end{align}
and the time-averaged Hamiltonian is  
\begin{align}
	 H_0 = \frac{p^2}{2m} + \frac{1}{2}m\omega_0^2 x^2 + \frac{1}{4}m \omega_0^2A^2.
\end{align}
The Fourier components of the potential are 
\begin{align}
	V^{(0)} & = \frac{1}{2}m\omega_0^2 x^2 + \frac{1}{4}m\omega_0^2A^2,\\
	V^{(1)}& = -\frac{1}{2}i A x m \omega_0^2 = -V^{(-1)}, \\
	V^{(2)} &= -\frac{1}{8}A^2 m \omega_0^2 = V^{(-2)}.
\end{align}
We immediately see that $H_{\text{eff}}^{(1)} = 0$, since $|V^{(-j)}|=|V^{(j)}|$.
Computing the commutators, we find that the second-order term is  
\begin{align}
	H_\text{eff}^{(2)} & = \frac{A^2 m \omega_0^4}{4 \Omega^2}.
\end{align}
Thus, to second order, we have 
\begin{align}
	H_{\text{eff}} &= \frac{p^2}{2m} + \frac{1}{2}m\omega_0^2 x^2 + \frac{1}{4}m\omega_0^2A^2 +  \frac{A^2 m \omega_0^4}{4 \Omega^2},
\end{align}
which gives a quasienergy spectrum 
\begin{align}
	\varepsilon_\alpha & = \left( \alpha + \frac{1}{2} \right)\hbar \omega_0 + \frac{1}{4} m\omega_0^2 A^2 + \frac{A^2 m \omega_0^4}{4 \Omega^2}	.
\end{align}

The time-averaged energy can be given in terms of the quasienergy by using the Hellmann-Feynman theorem \cite{Zeldovich1967,Grifoni1998c}.
\begin{align}
       \bar E_\alpha = \varepsilon_\alpha - \Omega \p{\varepsilon_\alpha}{\Omega},
\end{align}
and we finally obtain the expression for the time-averaged energy to second order\begin{align}
	\bar E_{\alpha} & =  \left( \alpha + \frac{1}{2} \right)\hbar \omega_0 + \frac{1}{4} m \omega_0^2 A^2 + \frac{A^2 m \omega_0^4}{2 \Omega^2}	.
\end{align}

\bibliography{library}{}
\end{document}